\documentclass[a4]{aa}
\usepackage{graphicx}
\usepackage{times}

\def \ref {\noindent\hangindent=1.0in\hangafter=1}

\def\ltsima{$\; \buildrel < \over \sim \;$}
\def\simlt{\lower.5ex\hbox{\ltsima}} % < over ~
\def\gtsima{$\; \buildrel > \over \sim \;$}
\def\simgt{\lower.5ex\hbox{\gtsima}} % > over ~
\def \integral{\emph{INTEGRAL}}

\begin{document}

\title{\integral\ observations of the field of the BL Lacertae object 
S5~0716+714\thanks{Based on observations obtained 
with \integral, an ESA project with instruments and science
data center funded by ESA member states (especially the PI countries:
Denmark, France, Germany, Italy, Switzerland, Spain, Czech Republic and
Poland), and with the participation of Russia and the USA.}}

\authorrunning{E. Pian et al.} 
\titlerunning{S5~0716+714 with \integral} 

\author{E.~Pian\inst{1,2}, 
L.~Foschini\inst{2},
V.~Beckmann\inst{3,4},
A.~Sillanp\"a\"a\inst{5},
S.~Soldi\inst{6},
G.~Tagliaferri\inst{7}, 
L.~Takalo\inst{5},
P.~Barr\inst{8},
G.~Ghisellini\inst{7},
G.~Malaguti\inst{2},
L.~Maraschi\inst{9},
G.G.C.~Palumbo\inst{10},
A.~Treves\inst{11},
T.~J.-L. Courvoisier\inst{6,12},
G.~Di Cocco\inst{2},
N.~Gehrels\inst{3},
P.~Giommi\inst{13},
R.~Hudec\inst{14},
E.~Lindfors\inst{5},
A.~Marcowith\inst{15},
K.~Nilsson\inst{5},
M.~Pasanen\inst{5},
T.~Pursimo\inst{16},
C.M.~Raiteri\inst{17},
T.~Savolainen\inst{5},
M.~Sikora\inst{18}, 
M.~Tornikoski\inst{19},
G.~Tosti\inst{20},
M.~T\"urler\inst{6},
E.~Valtaoja\inst{5},
M.~Villata\inst{17},
R.~Walter\inst{6,12}
}

\institute{INAF, Osservatorio Astronomico di Trieste, Via G.B. Tiepolo 11, 
I-34131 Trieste, Italy
\and
IASF-CNR, Sezione di Bologna, Via Gobetti 101, I-40129 Bologna, Italy
\and
NASA Goddard Space Flight Center, Code 661, Greenbelt, MD 20771, USA
\and
Joint Center for Astrophysics, Department of Physics, University of Maryland, 
Baltimore County, MD 21250, USA
\and
Tuorla Observatory, University of Turku, 21500 Piikki\"o, Finland
\and
\integral\ Science Data Center, Chemin d'\'Ecogia 16, 1290 Versoix, Switzerland
\and
INAF, Osservatorio Astronomico di Brera, Via E. Bianchi, 46, I-23807 Merate 
(LC), Italy
\and
ESA-ESTEC, RSSD, Keplerlaan 1, Postbus 299, 2200 AG Noordwijk, The Netherlands
\and
INAF, Osservatorio Astronomico di Brera, Via Brera 28, I-20121 Milano, Italy
\and
Department of Astronomy, University of Bologna, Via Ranzani 1, I-40127 
Bologna, Italy
\and
Dipartimento di Fisica e Matematica, University of Insubria, Via Valleggio 11, I-22100 
Como, Italy
\and
Observatoire de Gen\`eve, 51 Ch. des Maillettes, 1290 Sauverny, Switzerland 
\and
ASI Science Data Center, Via Galileo Galilei, I-00044 Frascati, Italy
\and
Astronomical Institute, CZ-251 65 Ondrejov, Czech Republic
\and
Centre d'\'Etude Spatiale de Rayonnements, 31028 Toulouse, France
\and
Nordic Optical Telescope, Apartado 474, E-38700 Santa Cruz de La Palma, Spain
\and
INAF, Osservatorio Astronomico di Torino, Via Osservatorio 20, I-10025 Pino 
Torinese (TO), Italy
\and
Nikolaus Copernicus Astronomical Center, Bartycka 18, 00-716 Warsaw, Poland
\and
Mets\"ahovi Radio Observatory, Mets\"ahovintie 114, FIN-02540 Kylm\"al\"a, 
Finland
\and
Department of Physics, University of Perugia, Via A. Pascoli, I-06123 Perugia,
Italy
}

\offprints{E. Pian \\ \email{pian@ts.astro.it}}   

\abstract{We have performed observations of the blazar S5~0716+714
with \integral\ on 2-6 April 2004.  In the first months of 2004, the
source had increased steadily in optical brightness and had undergone
two outbursts.  During the latter, occurred in March, it reached the
extreme level of $R = 12.1$ mag, which triggered our \integral\
program. The target has been detected with IBIS/ISGRI up to 60 keV,
with a flux of $\sim 3 \times 10^{-11}$ erg s$^{-1}$ cm$^{-2}$ in the
30-60 keV interval, a factor of $\sim$2 higher than observed by the
{\it BeppoSAX} PDS in October 2000. In the field of S5~0716+714 we
have also detected the Flat Spectrum Radio Quasar S5~0836+710 and the
two Seyfert galaxies Mkn~3 and Mkn~6. Their IBIS/ISGRI spectra are
rather flat, albeit consistent with those measured by {\it BeppoSAX}.  
In the spectrum of Mkn~3 we find some evidence of a
break between $\sim$60 and $\sim$100 keV, reminiscent of the high
energy cut-offs observed in other Seyfert galaxies.  
This is the first report of \integral\ spectra of weak Active
Galactic Nuclei. 
}

\maketitle 

\section{Introduction}

The high  energy emission of  Active Galactic Nuclei  (AGN) 
carries the most direct and constraining information on the radiation
mechanisms and the nature of  the  central  engine.    Blazar  type  AGNs,
traditionally subdivided in Flat  Spectrum Radio Quasars (FSRQ) and BL
Lacertae  Objects  (\cite{cmu1995}),  are  very powerful  and
variable  multiwavelength emitters.

At gamma-ray  wavelengths, their spectral output has often a maximum
and their variability
exhibits the largest amplitudes  (\cite{mhu1997}), making
them suitable  targets for the \integral\ mission, 
particularly during active states. The
radio-quiet  and less  luminous  Seyferts exhibit  hard X-ray  spectra
often  extending to  the  soft gamma-ray  domain.   At these  energies
spectral cut-offs have been detected in many of these objects by soft 
gamma-ray experiments like the {\it CGRO} OSSE and the 
{\it BeppoSAX} PDS. Studying these spectral features
is  relevant  to  the  identification of  the emission
mechanism  responsible  for the production of the spectrum at those energies 
(e.g., \cite{rs96}; \cite{hmg97}; \cite{pop00}).

The BL Lac object S5~0716+714 has been monitored at radio and optical 
wavelengths by more than 40 telescopes in the northern hemisphere during 
a Whole Earth Blazar Telescope
(\cite{mv2004}) campaign lasting from September 2003 to June 2004. 
The source had been already observed by \integral\  
in October 2003 during an optically active state (Wagner et al. 2004, in prep.).
In January and 
March 2004, S5~0716+714 was in outburst and achieved its optical historical  maximum.
In late March 2004 it brightened by  2 magnitudes with respect to  its November 2003
level, and  by 1 magnitude in  $\sim$2 weeks, reaching  a magnitude of
12.1  in  the  optical  $R$-band (Sillanp\"a\"a et  al. 2004, in  prep.).   

The large  optical variation
observed in March  2004 matched the trigger criteria  for our \integral\
Target-of-Opportunity  program for blazars  in outburst  (Proposal ID:
220049),  thus  we  activated  the campaign.   Observations with {\it RXTE} 
and
{\it XMM-Newton} as well as with ground-based optical and radio
telescopes have been  carried  out simultaneously  with
\integral.  The results of this multiwavelength monitoring will be 
reported in forthcoming papers.
We  report  here  on  the   \integral\  observations  of  the  field  of
S5~0716+714, in  which we  have also detected  another blazar  and two
Seyfert galaxies.

\section{Observations, data analysis and results}
\label{}

\integral\ (\cite{cw2003}) observed S5~0716+714 (Galactic coordinates:
$l = 144^{\circ}$, $b = +28^{\circ}$) starting 
from 2004 April $2^{\rm nd}$, 20:49:25, and  ending on 
2004 April $7^{\rm th}$, 00:14:08 UT. In order to optimize the  performance of the
SPI spectrometer (\cite{gv2003}) we adopted an observing scheme
consisting in  one pointing  of about  $2$ ks, followed  by a  slew of
$120$  s,  then  by  another  pointing,  and so  on,  as  to  build  a
rectangular pattern (dither pattern $5\times 5$).

The total duration of the observation was of $280$ ks, but the effective
exposures of the IBIS/ISGRI (\cite{pu2003}; \cite{lll03}), 
IBIS/PICsIT (\cite{pu2003}; \cite{gdc03}),
SPI (\cite{gv2003}) and JEM-X (\cite{nl2003}) detectors were of
256, 237, 218, and 189 ks, 
respectively (only JEM-X1 was used, while JEM-X2 was switched off). 
This reduction  is  due to  telemetry gaps and dead time corrections (generally
affecting  every \integral\  observation), to  the  occurrence of  a
failure of the  VETO module n. 15,  that caused IBIS to be  idle for 9 ks,
and to the removal of 3
pointings with 13 ks from the SPI data set because of problems in fitting
them.

The  screening, reduction,  and analysis  of the \integral\ data
have been  performed by  using the \integral\  
Offline Scientific
Analysis  (OSA) V.~4.0, 
publicly  available through  the \integral\ Science Data
Center\footnote{http://isdc.unige.ch/index.cgi?Soft+download}
(ISDC, \cite{tjlc2003a}).  The algorithms implemented  in the
software  are   described  in  Goldwurm  et  al.    (2003)  for  IBIS,
Westergaard et al. (2003) for JEM-X,  and Diehl et al. (2003) for SPI,
and we refer  the reader to these papers for more  details on the data
processing  and  deconvolution  of  coded--mask  telescopes  on  board
\integral.

Only the  observations with  the IBIS/ISGRI instrument  yielded 
significant source
detections, as reported in Sections 2.1 and 2.2.  The IBIS/PICsIT, SPI
and JEM-X data  were also accumulated into final  images.  However, no
sources are detected in those  data.  

The upper limit for IBIS/PICsIT
is $7.3 \times 10^{-10}$ erg s$^{-1}$ cm$^{-2}$ in the
252--336 keV range (292 mCrab). 
In the
most sensitive SPI energy range, 20--40 keV, the spectrometer achieved a 
marginally significant detection for Mkn~3: 
$8.3 \pm 3.5 \times 10^{-4}$ ph s$^{-1}$ cm$^{-2}$ (4.6 mCrab). For 
S5~0716+714, S5~0836+710, and Mkn~6 
the 3$\sigma$ upper limit is 
$\sim 10^{-3}$ ph s$^{-1}$ cm$^{-2}$ (6 mCrab).
The extrapolation of the ISGRI spectra of S5~0836+710, Mkn~3 and Mkn~6
(see Sections 2.1 and 2.2) to lower energies falls a factor from 2 to 20 
below  the sensitivity  of the  JEM-X image.  The JEM-X
3$\sigma$ upper  limits for 
S5~0716+714, S5~0836+710, Mkn~3 and Mkn~6 are 6, 10, 8, 6 mCrab (5-20 keV), 
respectively.

No data were acquired with the Optical Monitor (\cite{jmm2003}).

\subsection{IBIS observation of S5~0716+714}

A first  inspection of  the IBIS/ISGRI data  revealed the  presence of
high  background, with some structures.  However,  
in the final  mosaic of  all the
available data (i.e., the weighted combination of the individual pointings),
no systematic  effects have been  found.

S5~0716+714 is detected  with signal-to-noise 
ratio $4.5 \sigma$ in the energy band $30-60$ keV, for a count rate of 
$0.11 \pm 0.04$ counts s$^{-1}$. 
Since the source was  better detected in the first part of the \integral\ 
observation, indicating that it was declining,
we selected and accumulated 
the individual pointings of the early portion of the monitoring, for  
which  the signal-to-noise ratio at the 
position of the blazar is larger than 1.  
This reduces the useful  exposure to a total of 84 ks, but allows us to 
improve the significance of the detection of S5~0716+714 to $6.5\sigma$ 
in the $30-60$~keV energy range.  
No signal is detected above 60 keV. 

Given  the  low  signal-to-noise  ratio, 
it  was  not possible to study the intra-orbit variability of the source, 
nor to extract a spectrum.  

In order to evaluate the flux of the source we assumed
a spectrum identical to that of the Crab nebula ($\Gamma = 2.1$) and scaled 
the normalization via the count rates ratio in the 30-60 keV range (Table~1).
Note that
our flux measurement is only representative of
the higher state of the source during the observation, and not of the overall
average state.

\subsection{Other AGNs in the IBIS field of view} 

The large field of view of IBIS ($19^{\circ}\times 19^{\circ}$ at half
response)    allowed    us    to   observe    serendipitously    other
sources.  Specifically,  three additional  AGNs  are  detected in  the
IBIS/ISGRI  image  with higher  significance  than  our blazar  target
(Table~1), and up to 100  keV.  Also for these sources, the detections
in the individual pointings are not  significant,  and therefore  it is  not
possible to  study their intra-orbit variability. However,  the 
signal-to-noise ratio of
the summed  image is sufficiently high  to allow the  extraction of an
average spectrum.  

Since spectral reconstruction is very sensitive to the background correction in weak
sources, we have extracted the spectra following two independent procedures,
as recommended in the OSA 
guidelines\footnote{http://isdc.unige.ch/Soft/download/osa/osa\_sw/osa\_sw-4.0/osa\_issues-4.0.txt}.
For each source, we have obtained both a coadded spectrum from the spectra extracted from
individual pointings, and a spectrum constructed from the imaging in 
the same energy bands used for spectral extraction.
These are found to be consistent with each other in all sources. We have used the spectra
obtained with the former method for spectral analysis.
In order to apply the $\chi^2$ statistics, the 
spectral signal has been binned in intervals where
the significance is at least $3 \sigma$.

In the energy range $\approx 60-80$~keV the IBIS spectra are affected by 
instrumental features (\cite{terr03}), 
that are  not well  
modeled by  
the response  matrix. The OSA V.~4 software reduces this problem only partially, 
therefore, considering also the relevance of the
background relative to the flux levels of our sources, 
we conservatively excluded the above spectral region from the analysis.

We fitted single power-laws to the spectra of 
S5~0836+710 (Fig.~1) 
and 
Mkn~6 (Fig.~2)  
using
the  \texttt{xspec} package  (v.  11.3.1).   
No systematic errors have been added in the fitting; we estimate the flux
calibration uncertainties to be of the order of $\sim$10\%.

In  Mkn~3, the  spectral 
point  at
$\sim$100 keV  (see Fig.~3)  makes a single power-law  fit inadequate
(the $\chi^2$ is 13 for 4 degrees of freedom);
therefore we used  a broken power-law, although the  high energy index
$\Gamma_2$ is  obviously poorly constrained  by the isolated  point at
100  keV (see contour plot in Fig.~4).  

We have also tried a fit with a power-law plus an exponential cutoff. If the power-law 
photon index is left as a free 
parameter, it assumes an unreasonably flat value, $\Gamma \simeq -0.07$, inconsistent with the 
power-law index which best fits 
the spectral points at energies lower than 100 keV, $\Gamma = 1.3$. 
A contour plot of the fitted photon index vs  high energy cut-off is shown in Fig.~5.
By freezing instead
the photon index to $\Gamma = 1.3$, we obtain  a cut-off energy of $\sim$87~keV with an
unacceptably high 
$\chi^2$ of 8.6 for 4 degrees of freedom.  By adding in quadrature to the statistical 
uncertainties a systematic error as large as 25\% we recover a $\chi^2 = 4$ for 4 degrees
of freedom and the cut-off energy is $E = 95^{+105}_{-40}$~keV. 

Therefore, we
formally prefer the broken power-law rather than the cut-off power-law model.  
At best, the latter suggests a cut-off energy larger than $\sim$50 keV.
We stress that the choice of the best interpretation of the spectral shape of Mkn~3 largely
relies upon the point at $\sim$100 keV, that is very sensitive  to      background 
subtraction. A more robust spectral modeling must await a more accurate spectral
measurement at these energies.

The results  of  the  spectral  analysis are  reported  in
Table~1.  The spectra, along  with their best-fit models, are reported
in Figures~1,2,3.

% --------------     Table 1: The Sources    -----------------

\begin{table*}[t!] 
\caption[]{Sources detected in the IBIS/ISGRI field of S5~0716+714}
\begin{center}
\begin{tabular}{lccccccccccc}
\noalign{\smallskip}       
\hline
\noalign{\smallskip}
Object & AGN Type & $z$ & CR$^a$ & Flux$^b$ & Range$^c$ &  
$\Gamma_1^d$ & $\Gamma_2^d$ & $E_b^e$ & $\chi^2$ & d.o.f. & $\Gamma_{PDS}^f$ \\
\noalign{\smallskip}
\hline
S5~0716+714 & BL Lac & ...   & $0.36 \pm 0.07^g$ & 3.1 & 30-60 & ... & ... & ...  & ...  & ... & $1.6 \div 2.0$ \\
S5~0836+710 & FSRQ & 2.172 & $0.54 \pm 0.05$ & 4.6 & 20-100 & $1.3 \pm 0.3^h$ & ... & ... & 4.1 & 5 & $1.31 \pm 0.03$ \\  % flux30-60=1.8
Mkn~6 & Sy 1.5 & 0.019 & $0.49 \pm 0.06$ & 4.6 & 20-100 & $1.5^{+0.5}_{-0.4}$ & ... & ... & 3.7 & 3 & $1.8 \pm 0.2$ \\ % flux30-60=1.9
Mkn~3 & Sy 2 & 0.013 & $0.82 \pm 0.05$ & 7.4 & 20-100 & 1.3$^i$ & $>2.5$ & $80 \pm 20^h$ & 2.5 & 3 & $1.8 \pm 0.1$ \\ %f30-60=3.1
\noalign{\smallskip}         
\noalign{\smallskip}
\hline
\noalign{\smallskip}
\multicolumn{12}{l}{$^a$ IBIS/ISGRI count rate in the detection energy
range (Col.~6), in counts s$^{-1}$.}\\
\multicolumn{12}{l}{$^b$ Fitted flux in the detection energy range (Col.~6),
in $10^{-11}$~erg~s$^{-1}$~cm$^{-2}$. Calibration uncertainties are $\sim$10\%.}\\
\multicolumn{12}{l}{$^c$ Energy interval to which count rates (Col.~4) 
and fluxes (Col.~5) are referred, in keV.}\\
\multicolumn{12}{l}{$^d$ Photon index: $f_E \propto E^{-\Gamma}$.}\\
\multicolumn{12}{l}{$^e$ Break energy, in keV.}\\
\multicolumn{12}{l}{$^f$ Photon index measured with the {\it BeppoSAX} PDS
at previous epochs.}\\
\multicolumn{12}{l}{$^g$ Uncertainties on the count rates are 1$\sigma$.}\\
\multicolumn{12}{l}{$^h$ Uncertainties on the photon indices and break energy are
1.6$\sigma$.}\\
\multicolumn{12}{l}{$^i$ $\Gamma_1$ has been frozen to the index
of the single power-law which best
fits the spectrum below 100 keV, $\Gamma = 1.3 \pm 0.4$.}\\
\end{tabular}
\end{center}
\end{table*}                              

% --------------     Table 1: End     -----------------

%\begin{figure*}
%\begin{center}
%\psfig{file=pks0716_fig1.ps,width=8.5cm}
%\centerline{ 
%\includegraphics[scale=0.5]{new_mosaic_30_60_all.ps}
%\includegraphics[scale=0.5]{new_mosaic_30_60_0716.ps}       
%}
%\caption{{\it Left panel:} IBIS/ISGRI mosaic in 30-60 keV of the field 
%of S5~0716+714. {\it Right panel:} enlargement of the area of S5~0716+714. The
%mosaic has been constructed using the pointings in which the
%signal-to-noise ratio is larger than 1}
%\end{center}
%\end{figure*}

\begin{figure}
\centering
\includegraphics[angle=270,scale=0.3]{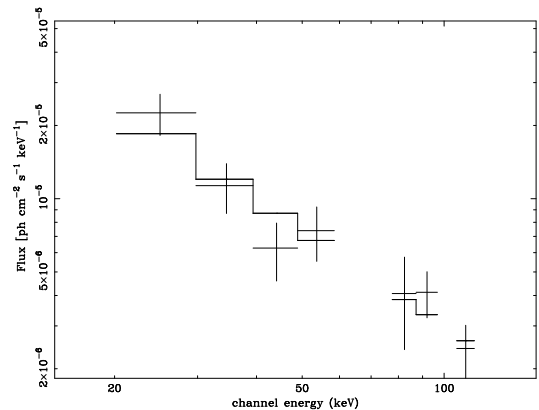}
\caption{IBIS/ISGRI spectrum of S5~0836+710. The overplotted step-like 
curve is the best fit single power-law}
\end{figure}

\begin{figure}
\centering
\includegraphics[angle=270,scale=0.3]{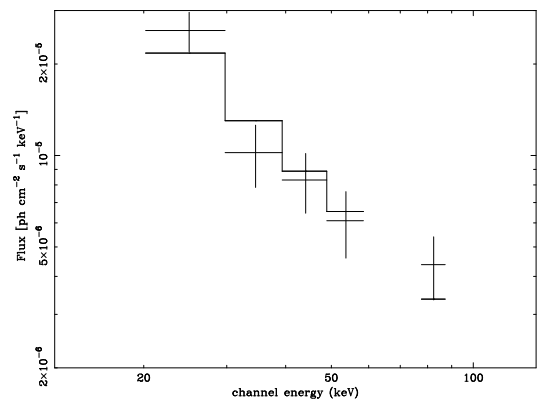}
\caption{IBIS/ISGRI spectrum of Mkn~6. The overplotted step-like curve 
is the best fit single power-law}
\end{figure}

\begin{figure}
\centering
\includegraphics[angle=270,scale=0.3]{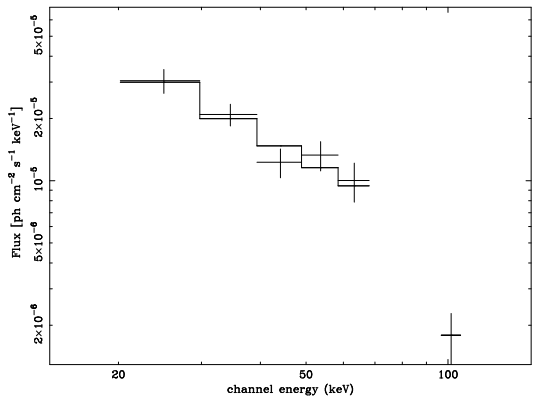}
\caption{IBIS/ISGRI spectrum of Mkn~3. The overplotted step-like curve 
is the single power-law which best fits the spectrum below 100 keV. The point
at 100 keV is clearly below this curve}
\end{figure}

\begin{figure}
\centering
\includegraphics[angle=270,scale=0.3]{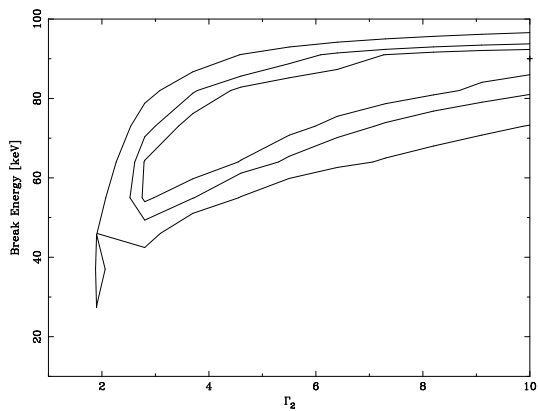}
\caption{Contour plot of  break energy vs photon index for the broken power-law model
fit of Mkn~3 spectrum}
\end{figure}

\begin{figure}
\centering
\includegraphics[angle=270,scale=0.3]{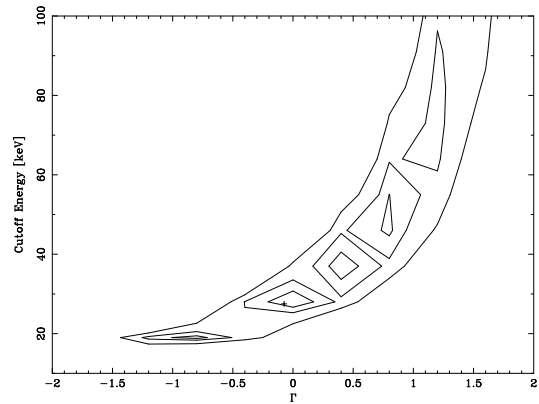}
\caption{Contour plot of cut-off energy vs photon index for the cut-off power-law model
fit of Mkn~3 spectrum}
\end{figure}

\section{Discussion}
\label{}

We  observed  the  blazar  S5~0716+714  with \integral\  while  it  was
undergoing  a major  optical outburst  and  detected the  source in  a
somewhat higher (about a factor of 2) gamma-ray state 
than observed in
October 2000  (\cite{gt2003}).  At  that epoch  the optical
flux  was slightly lower ($R \simeq 12.5$) than that observed  in  
March 2004 at maximum brightness.  The {\it BeppoSAX} PDS spectrum
suggests that the soft gamma-rays are due to inverse Compton scattering
of relativistic particles (\cite{gt2003}; \cite{giommi99}; see
also last column of Table~1, where the indices of the PDS spectra of our
sources are reported for comparison with the present results). 

In the  same field  of our  primary target we  also detected  the high
redshift blazar  S5~0836+710, belonging to the 
FSRQ sub-class (\cite{cmu1995}), and two bright Seyfert
galaxies, Mkn~3 and Mkn~6, all with brighter fluxes than S5~0716+714
(Table~1).   

The spectral index of S5~0836+710 is consistent with
that determined  through  {\it CGRO} BATSE  (\cite{am2000}) and 
{\it BeppoSAX} PDS (\cite{ft2000}) observations during
higher emission states (our measured flux is a factor of $\sim$3 lower than
found by both BATSE and {\it BeppoSAX}).  The flat spectral slope
favours the interpretation of the high energy spectrum as Compton scattering
of relativistic electrons off external radiation, as opposed to
synchrotron-self 
Compton, in FSRQ (\cite{sbr94}).

The IBIS/ISGRI spectral slope of  the Seyfert Mkn~6 is flatter, although
consistent with
that measured by  the {\it BeppoSAX} PDS (\cite{am2003a}; \cite{immler03}), 
within our rather
large errors. Similarly, the flux level in the IBIS/ISGRI observation is 
consistent with that
measured by the PDS.

Our IBIS/ISGRI spectrum of Mkn~3 has an index consistent with that determined by 
OSSE ($\Gamma \simeq 1.3$, \cite{zpj2000}) and
suggests the presence of a high energy
cut-off, at an energy approximately 
compatible with those of the breaks detected in other 
Seyfert objects and
radiogalaxies by OSSE and {\it BeppoSAX}, although on the lower energy 
side with respect to the average break energy determined by {\it BeppoSAX}
for Seyfert 2 galaxies
(\cite{jwn97}; \cite{ghg98}; \cite{lp98}; \cite{fn2000}; \cite{zg2001};
\cite{zpj2000}; \cite{gondoin01}; \cite{adr02}; \cite{am2003b}; \cite{dc03}). 
However, no evidence of a cut-off was found in Mkn~3 by Cappi
et al.  (1999) at
least up to 150 keV, the maximum energy at which the source was detected by
the {\it BeppoSAX} PDS.  Therefore, while considering our 
detection of a cut-off tentative, we cannot exclude a variable 
cut-off energy. We note
in fact that the flux detected by the PDS was $\sim$30\% higher
than that detected by IBIS, and the spectral slope ($\Gamma \simeq 1.8$) 
marginally steeper (see \cite{cappi1999}).  A  similar correlation of 
cut-off energy and spectral steepness has been observed with {\it BeppoSAX} 
in the Seyfert 1 NGC~5548 (\cite{fn2000}; \cite{pop00}). 
A better exposed \integral\ ISGRI and/or SPI spectrum would be necessary to 
confirm our finding.

\integral\ has significantly detected so far a number of radio-quiet and
radio-loud AGNs  (\cite{lb2004}; \cite{vb2004}; \cite{ajb2004}), among  
which the  brightest known  radio-loud  AGN, 3C~273
(\cite{tjlc2003b}).   Our  observation  of S5~0716+714  and
serendipitous detection of three additional AGNs proves \integral\ to be
effective, even  with relatively  short exposure observations,  in the
study of  bright extragalactic sources at high  Galactic latitudes and
underscores the importance of instruments with a large field of view and
good angular  resolution for the investigation of  gamma-ray-loud AGNs.
S5~0836+710 is  the second highest redshift AGN  detected by \integral,
after  PKS~1830-21  at  $z  =  2.507$ (\cite{lb2004}),  which
indicates that \integral\ can play  a role also in exploring
the high redshift universe.

\begin{acknowledgements}

We  thank the  staff at  the \integral\  Science Operation  Center and
Science  Data Center  for  their assistance  in  data acquisition  and
analysis,  and   in  particular  A.~Gros,   P.~Kretschmar,  A.~Parmar,
K.~Pottschmidt, R.~Williams.  An anonymous referee is acknowledged for
a constructive  report.  We acknowledge the EC  funding under contract
HPRCN-CT-2002-00321  (ENIGMA  network).  EP  and  LF  acknowledge  the
Italian Space Agency (ASI) for financial support.

\end{acknowledgements}

%\newpage


\begin{thebibliography}{}{}

\bibitem[Bassani  et  al.  2004]{lb2004}  Bassani,  L.,  Malizia,  A.,
Stephen, J.B., et al.  2004, Proceedings of the V \integral\ Workshop,
Munich 16-20 February 2004.  ESA SP-552, in press (astro-ph/0404442)
      
\bibitem[Beckmann et al. 2004]{vb2004} Beckmann, V., Gehrels, N., Favre, P., et al.  2004,
ApJ, in press (astro-ph/0406553)

\bibitem[Bird et al. 2004]{ajb2004}  Bird,   A.J.,  Barlow,  E.J.,   Bassani,  L.,  et
al. 2004, ApJ, 607, L33

\bibitem[Cappi et al. 1999]{cappi1999} Cappi, M., Bassani, L., Comastri, A., et al. 1999,
A\&A, 344, 857

\bibitem[Courvoisier et al. 2003a]{tjlc2003a} Courvoisier, T. J.-L., Walter, R., 
Beckmann, V., et al. 2003a, A\&A, 411, L53

\bibitem[Courvoisier et al. 2003]{tjlc2003b} Courvoisier, T. J.-L., Beckmann, V., 
Bourban, G., et al. 2003b, A\&A, 411, L343

\bibitem[Deluit \& Courvoisier 2003]{dc03} Deluit, S. \& Courvoisier, T. J.-L. 2003, A\&A, 399, 77

\bibitem[De Rosa et al. 2002]{adr02} De Rosa, A., Piro, L., Fiore, F., et al. 2002, A\&A, 387, 838

\bibitem[Di Cocco et al. 2003]{gdc03} Di Cocco, G., Caroli, E., Celesti, E. et al. 2003, 
A\&A, 411, L189
 
\bibitem[Diehl et al. 2003]{diehl} Diehl, R., Baby, N., Beckmann, V., et al. 2003, A\&A, 
411, L117

\bibitem[Giommi et al. 1999]{giommi99} Giommi, P., Massaro, E., Chiappetti, L., et al., 1999,
A\&A, 351, 59

\bibitem[Goldwurm et al. 2003]{goldwurm1} Goldwurm, A., David, P., Foschini, L., et al., 2003, 
A\&A 411, L223

\bibitem[Gondoin 2001]{gondoin01} Gondoin, P., Barr, P., Lumb, D., et al. 
2001, A\&A, 378, 806

\bibitem[Grandi et al. 1998]{ghg98} Grandi, P., Haardt, F., Ghisellini, G., et al. 
1998, ApJ, 498, 220

\bibitem[Haardt et al. 1997]{hmg97} Haardt, F., Maraschi, L., \& Ghisellini, G. 1997, ApJ, 
476, 620

\bibitem[Immler et al. 2003]{immler03} Immler, S., Brandt, W.N., Vignali, C., et al. 
2003, AJ, 126, 153

\bibitem[Johnson et al. 1997]{jwn97} Johnson, W.N., McNaron-Brown, K., Kurfess, J.D., et al. 
1997, ApJ, 482, 173

\bibitem[Lebrun et al. 2003]{lll03} Lebrun, F., Leray, J.P., Lavocat, P., et al., 2003,
A\&A 411, L141

\bibitem[Lund et al. 2003]{nl2003} Lund, N., Budtz-J\o rgensen, G., 
Westergaard, N.J., et al. 2003, A\&A, 411, L231

\bibitem[Malizia et al. 2000]{am2000} Malizia, A., Bassani, L., Dean, A.J., et al.
2000, ApJ, 531, 642

\bibitem[Malizia et al. 2003a]{am2003a} Malizia, A., Bassani, L., Capalbi, M., et al.
2003a, A\&A, 406, 105

\bibitem[Malizia et al. 2003b]{am2003b} Malizia, A., Bassani, L., Stephen, J.B., 
et al. 2003b, ApJ, 589, L17

\bibitem[Mas-Hesse et al. 2003]{jmm2003} Mas-Hesse, J.M., Gim\'enez, A., 
Culhane, L., et al. 2003, A\&A, 411, L261

\bibitem[Nicastro et al. 2000]{fn2000} Nicastro, F., Piro, L., De Rosa, A., et al. 2000, 
ApJ, 536, 718

\bibitem[Petrucci et al. 2000]{pop00} Petrucci, P.O., Haardt, F., Maraschi, L., et al. 
2000, ApJ, 540, 131

\bibitem[Piro et al. 1998]{lp98} Piro, L., Nicastro, F., Feroci, M., et al. 1998, 
Nuclear Physics B (Proc. Suppl.), 69/1-3, 481

\bibitem[Sikora et al. 1994]{sbr94} Sikora, M., Begelman, M.C., \& Rees, M.J. 1994, ApJ, 421, 153

\bibitem[Svensson 1996]{rs96} Svensson, R. 1996, ApJS, 92, 585

\bibitem[Tagliaferri et al. 2003]{gt2003} Tagliaferri, G., Ravasio, M., Ghisellini, G., et
al. 2003, A\&A, 400, 477

\bibitem[Tavecchio et al. 2002]{ft2000} Tavecchio, F., Maraschi, L., Ghisellini, G., et al.
2000, ApJ, 543, 535

\bibitem[Terrier et al. 2003]{terr03} Terrier, R., Lebrun, F., Bazzano, A., et al., 2003, A\&A, 411, L167

\bibitem[Ubertini et al. 2003]{pu2003} Ubertini, P., Lebrun, F., Di Cocco, G., et al. 2003,
A\&A, 411, L131

\bibitem[Ulrich et al. 1997]{mhu1997} Ulrich, M.-H., Maraschi, L., Urry, C.M. 1997, ARA\&A,
35, 445

\bibitem[Urry \& Padovani 1995]{cmu1995} Urry, C.M., \& Padovani, P. 1995, PASP, 107, 803

\bibitem[Vedrenne et al. 2003]{gv2003} Vedrenne, G., Roques, J.-P., Sch\"onfelder, V., et
al. 2003, A\&A, 411, L63

\bibitem[Villata et al. 2004]{mv2004} Villata, M., Raiteri, C.M., Kurtanidze, O.M., 
et al., 2004, A\&A, 421, 103

\bibitem[Westergaard et al. 2003]{west} Westergaard, N.J, Kretschmar, P., Oxborrow, C.A., et al. 2003,
A\&A, 411, L257

\bibitem[Winkler et al. 2003]{cw2003} Winkler, C., Courvoisier, T.J.-L., Di Cocco, G., et
al. 2003, A\&A, 411, L1

\bibitem[Zdziarski \& Grandi 2001]{zg2001} Zdziarski, A.A., \& Grandi, P. 2001, ApJ, 551, 186

\bibitem[Zdziarski et al. 2000]{zpj2000} Zdziarski, A.A., Poutanen, J., \& Johnson, N.W. 2000, 
ApJ, 542, 703

\end{thebibliography}
\end{document}